\documentclass[sigconf,natbib=true]{acmart}
\usepackage{graphicx} 
\usepackage{caption}
\usepackage[linesnumbered,ruled,vlined]{algorithm2e}
\usepackage{subfigure}
\usepackage{multirow}
\usepackage[normalem]{ulem}
\usepackage{balance}
\usepackage[T1]{fontenc}
\usepackage[utf8]{inputenc}
\usepackage{makecell}
\usepackage{enumitem}
\usepackage{amsmath}

\SetKwInput{KwInput}{Input}                
\SetKwInput{KwOutput}{Output}              

\AtBeginDocument{%
  \providecommand\BibTeX{{%
    \normalfont B\kern-0.5em{\scshape i\kern-0.25em b}\kern-0.8em\TeX}}}

\copyrightyear{2023}
\acmYear{2023}
\setcopyright{acmlicensed}\acmConference[SIGIR '23]{Proceedings of the 46th
International ACM SIGIR Conference on Research and Development in
Information Retrieval}{July 23--27, 2023}{Taipei, Taiwan}
\acmBooktitle{Proceedings of the 46th International ACM SIGIR Conference on
Research and Development in Information Retrieval (SIGIR '23), July 23--27,
2023, Taipei, Taiwan}
\acmPrice{15.00}
\acmDOI{10.1145/3539618.3592077}
\acmISBN{978-1-4503-9408-6/23/07}

\begin{document}

\title{Unbiased Pairwise Learning from Implicit Feedback for Recommender Systems without Biased Variance Control}


\author{Yi Ren}
\affiliation{%
  \institution{Tencent }
  \city{Beijing}
  \country{China}}  
\email{henrybjren@tencent.com}

\author{Hongyan Tang}
\affiliation{%
  \institution{Tencent}
  \city{Beijing}
  \country{China}}  
\email{violatang@tencent.com}

\author{Jiangpeng Rong}
\affiliation{%
  \institution{Tencent }
  \city{Beijing}
  \country{China}}    
\email{jprong124@gmail.com}

\author{Siwen Zhu}
\affiliation{%
  \institution{Tencent}
  \city{Beijing}
  \country{China}}  
\email{siwenzhu@tencent.com}

%
%
%

\renewcommand{\shortauthors}{Yi Ren, Hongyan Tang, Jiangpeng Rong, \& Siwen Zhu}

\begin{abstract}
Generally speaking, the model training for recommender systems can be based on two types of data, namely explicit feedback and implicit feedback. Moreover, because of its general availability, we see wide adoption of implicit feedback data, such as click signal. There are mainly two challenges for the application of implicit feedback. First, implicit data just includes positive feedback. Therefore, we are not sure whether the non-interacted items are really negative or positive but not displayed to the corresponding user. Moreover, the relevance of rare items is usually underestimated since much fewer positive feedback of rare items is collected compared with popular ones. To tackle such difficulties, both pointwise and pairwise solutions are proposed before for unbiased relevance learning. As pairwise learning suits well for the ranking tasks, the previously proposed unbiased pairwise learning algorithm already achieves state-of-the-art performance. Nonetheless, the existing unbiased pairwise learning method suffers from high variance. To get satisfactory performance, non-negative estimator is utilized for practical variance control but introduces additional bias. In this work, we propose an unbiased pairwise learning method, named \textbf{\textit{UPL}}, with much lower variance to learn a truly unbiased recommender model. Extensive offline experiments on real world datasets and online A/B testing demonstrate the superior performance of our proposed method.

\end{abstract}

\vspace{-0.2cm}

\vspace{-0.2cm}

\maketitle

\vspace{-0.1cm}
\section{Introduction}

Recommender systems usually rely on implicit user feedback for model training owning to the cheap cost of collecting such data \cite{liu2010personalized}. For this scenario, the typical model learning techniques \cite{koren2009matrix,rendle2012bpr,johnson2014logistic}, recognize interacted items as positive and all the other items as potential negative examples. There are mainly two difficulties to learn unbiased user preference based on implicit feedback data. First, we are not sure whether a non-interacted item is irrelevant to the user or positive but not yet recommended \cite{bekker2019beyond,elkan2008learning}. The second challenge is that the positive feedback is missing-not-at-random (MNAR) \cite{yang2018unbiased,steck2010training} since users are more likely to interact with the popular items because of the exposure bias and selection bias \cite{chen2020bias}.

To handle the aforementioned challenges, multiple methods \cite{hu2008collaborative,liang2016modeling,saito2020unbiased,saito2020unbiased2,lee2021dual} are proposed to learn unbiased models based on the Missing-Not-At-Random implicit feedback data.   
WMF \cite{hu2008collaborative} addresses the first problem of positive-unlabeled feedback by upweighting the training loss of the interacted ones. ExpoMF \cite{liang2016modeling} models the items' exposure probability for more accurate training weight of different items. Rel-MF \cite{saito2020unbiased} adopts inverse propensity weighting \cite{schnabel2016recommendations,swaminathan2015self,rosenbaum1983central} and derives the unbiased point-wise loss to achieve better recommendation quality. And MF-DU \cite{lee2021dual} separately estimate the exposure probability for interacted and non-interacted data for better bias correction. Unlike the point-wise algorithms of \cite{hu2008collaborative,liang2016modeling,saito2020unbiased,lee2021dual}, Unbiased Bayesian Personalized Ranking \cite{saito2020unbiased2} extends the pair-wise learning algorithm of \cite{rendle2012bpr} and formulates an unbiased objective function. As pairwise approach is more suitable than the pointwise methods for the ranking task of recommender systems \cite{song2018neural,wang2018collaborative}, UBPR \cite{saito2020unbiased2} achieves state-of-the-art recommendation performance. Nevertheless, UBPR \cite{saito2020unbiased2} suffers from high variance. As a result, non-negative estimator is applied to practically control the variance at the cost of introducing additional bias. 

One related but different line of research \cite{hu2019unbiased,joachims2017unbiased,ai2018unbiased,wang2018position,agarwal2019general,ren2022unbiased} focuses on learning unbiased learning to rank models by modelling position bias \cite{chen2020bias} as a counterfactual effect. In contrast, exposure bias and selection bias \cite{chen2020bias} are much more important in our setting. 

In this paper, we design an unbiased pairwise learning algorithm with lower variance so as to circumvent the step of variance control and perform truly unbiased learning. We summarize our main contributions below.
\vspace{-0.1cm}
\begin{itemize}[leftmargin=.2in]
\item We design an effective unbiased pairwise learning algorithm for implicit feedback with theoretically lower variance than the existing approach. 
\item Thanks to the proposed low variance algorithm, its unbiased pairwise loss function can be directly used without further bias-variance trade-off. 
\item We conduct offline experiments and online A/B testing to evaluate and understand the effectiveness of our method.
\end{itemize}
\vspace{-0.3cm}
\section{Preliminaries And Notations}
Let $u \in U$ represent a user and $i,j \in I$ denote an item. In addition, let $D_{point} = U \times I$ be the set of all possible interaction data. $c_{u,i}$ denotes the binary random variable of implicit interaction between $u$ and $i$. And $r_{u,i}$ is the binary random variable representing the relevance between $u$ and $i$. Furthermore, $o_{u,i}$ denotes the status of exposure $i$ to $u$. Only the interaction variables are observed for implicit feedback. The implicit feedback $c_{u,i}$ can be modeled below.
\vspace{-0.5cm}
\begin{align} 
P(c_{u,i}=1)&=P(o_{u,i}=1) \times P(r_{u,i}=1) \nonumber  \\
&=\theta_{u,i} \times \gamma_{u,i}, \forall(u,i) \in D_{point} 
\end{align}
\vspace{-0.5cm}

where $\theta_{u,i}=P(o_{u,i}=1)$ and $\gamma_{u,i}=P(r_{u,i}=1)$ represent the exposure and relevance probability respectively. Moreover, we use $\theta_{u,i}^-$ to represent the posterior exposure probability with negative implicit feedback and can be computed below.
\vspace{-0.3cm}

\begin{align} \label{eq:theta_minus}
&\!\!\! \!  \theta_{u,i}^-=P(o_{u,i}=1|c_{u,i}=0)=\frac { \theta_{u,i}  (1-\gamma_{u,i})}
 {\!  1 - \theta_{u,i}\gamma_{u,i}}
\end{align}
\vspace{-0.3cm}

And $P(r_{u,i} \!\!> \!\! r_{u,j})$, which equals $P(r_{u,i}=1,r_{u,j}=0)$, denotes the probability distribution that user $u$ prefers item $i$ over item $j$. Furthermore, given independent observation probability of different items, it is intuitive to derive $P(c_{u,i}=1, c_{u,j}=0, o_{u,i}=1, o_{u,j}=1)$ as follows.
\vspace{-0.3cm}
\begin{align} \label{eq:pair_relation}
&\!\!\! \!  P(c_{u,i}=1, c_{u,j}=0, o_{u,i}=1, o_{u,j}=1) \nonumber  \\
&\!\!\! \! \!\!=P(o_{u,i}=1)P(o_{u,j}=1)P(r_{u,i}=1,r_{u,j}=0)) \nonumber \\
&\!\!\! \! \!\!=\theta_{u,i} \theta_{u,j} P(r_{u,i}=1,r_{u,j}=0))
\end{align}
\vspace{-0.5cm}
\section{Methodology}
In this section, we propose an unbiased pairwise learning algorithm, named \textbf{\textit{UPL}}, with lower variance than the state-of-the-art methods. It performs truly unbiased pairwise learning based on implicit feedback by circumventing the step of external variance control. 
\vspace{-0.1cm}
\subsection{Existing Unbiased Pairwise Learning Method for Biased Implicit Feedback}
In the pairwise setting, the loss function is defined on the pair of positive item $i$ and negative item $j$ for user $u$. For this scenario, we let $f(u,i)$ denotes the ranker function to be learned to reflect the relevance between $u$ and $i$. And $L(f(u,i),f(u,j) )$ denotes the pairwise loss function. UBPR \cite{saito2020unbiased2} utilizes an unbiased estimator for the ideal pairwise loss and achieves the state-of-the-art results. The unbiased loss function is defined as follows, where $Reg(f)$ and $\lambda$ are the regularization loss and loss weight respectively. Unlike BPR \cite{rendle2012bpr} with strict pair defined on the implicit feedback, $c_{u,j}$ below can be positive or negative. 
\vspace{-0.1cm}
\begin{equation} 
 R_{ubpr}(f)  =    \sum_{u, i, j} { \frac { c_{u,i}}{\theta_{u,i}}  (1 - \frac { c_{u,j}}{\theta_{u,j}}) L(f(u,i),f(u,j))} + \lambda Reg(f)
\end{equation}
\vspace{-0.1cm}
Nonetheless, its variance depends on the inverse of the product of two propensity scores, namely $\theta_{u,i}$ and $\theta_{u,j}$. Thus, it tends to produce suboptimal results, especially for the tail items with low exposure probability. Furthermore, the unbiased estimator can take negative values by definition, thereby causing ever severe variance issues and hindering model convergence. Therefore, inspired by the work of positive-unlabeled learning \cite{kiryo2017positive}, non-negative estimator, which clips large negative values, is applied for practical variance control but introduces additional bias.
\begin{table*}
\centering
\begin{tabular}{lllll|lll|lll} 
\hline
\multirow{2}{*}{DataSets}  & \multirow{2}{*}{Methods} & \multicolumn{3}{c|}{\textbf{\textit{DCG@K}}}                                   & \multicolumn{3}{c|}{\textbf{\textit{Recall@K}}}                                & \multicolumn{3}{c}{\textbf{\textit{MAP@K}}}                                     \\ 
\cline{3-11}
                           &                          & \textit{K=3}             & \textit{K=5}             & \textit{K=8}             & \textit{K=3}             & \textit{K=5}             & \textit{K=8}             & \textit{K=3}             & \textit{K=5}             & \textit{K=8}              \\ 
\hline\hline
\multirow{7}{*}{Yahoo! R3} & \textit{WMF}                      & 0.08312                  & 0.09643                  & 0.10809                  & 0.08962                  & 0.11807                  & 0.15041                  & 0.06809                  & 0.07649                  & 0.08343                   \\
                           & \textit{Rel-MF}                  & 0.08586                  & 0.09894                  & 0.11019                  & 0.09243                  & 0.12040                  & 0.15165                  & 0.07052                  & 0.07896                  & 0.08574                   \\
                           & \textit{MF-DU}                    & 0.08493                  & 0.09828                  & 0.10967                  & 0.09150                  & 0.12002                  & 0.15160                  & 0.06995                  & 0.07854                  & 0.08538                   \\
                           & \textit{BPR}                      & 0.09077                  & 0.10415                  & 0.11405                  & 0.09745                  & 0.12577                  & 0.15327                  & 0.07455                  & 0.08313                  & 0.08917                   \\ 
\cline{2-11}
                           & \textit{UBPR}                     & 0.09418                  & 0.10745                  & 0.11693                  & 0.10127                  & 0.12941                  & 0.15555                  & 0.07792                  & 0.08667                  & 0.09263                   \\
                           & $UBPR_{NClip}$              & $0.09061^{-}$              & $0.10389^{-}$              & $0.11425^{-}$              & $0.09745^{-}$              & $0.12572^{-}$              & $0.15442^{-}$              & $0.07485^{-}$              & $0.08350^{-}$              & $0.08988^{-}$               \\ 
\cline{2-11}
                           & $UPL$                   & \textbf{\uline{0.09698}} & \textbf{\uline{0.11005}} & \textbf{\uline{0.11906}} & \textbf{\uline{0.10403}} & \textbf{\uline{0.13175}} & \textbf{\uline{0.15673}} & \textbf{\uline{0.08072}} & \textbf{\uline{0.08945}} & \textbf{\uline{0.09511}}  \\

\hline\hline
\multirow{7}{*}{Coat}      & $WMF$                      & 0.10320                  & 0.12415                  & 0.14580                  & 0.11322                  & 0.15815                  & 0.21844                  & 0.08616                  & 0.10017                  & 0.11392                   \\
                           & \textit{Rel-MF}                   & 0.10516                  & 0.12474                  & 0.14815                  & 0.11513                  & 0.15648                  & 0.22141                  & 0.08728                  & 0.10088                  & 0.11539                   \\
                           & \textit{MF-DU}                    & 0.10529                  & 0.12596                  & 0.14955                  & 0.11511                  & 0.15923                  & 0.22419                  & 0.08795                  & 0.10212                  & 0.11628                   \\
                           & $BPR$                      & 0.10048                  & 0.12127                  & 0.14451                  & 0.11023                  & 0.15477                  & 0.21932                  & 0.08357                  & 0.09717                  & 0.11158                   \\ 
\cline{2-11}
                           & $UBPR$                     & 0.10727                  & 0.12735                  & 0.14990                  & 0.11769                  & 0.16082                  & 0.22359                  & 0.08897                  & 0.10229                  & 0.11641                   \\
                           & $UBPR_{NClip}$              & 0.10597                  & 0.12649                  & 0.14900                  & $0.11598^{-}$              & 0.16013                  & 0.22279                  & 0.08808                  & 0.10142                  & 0.11520                   \\ 
\cline{2-11}
                           & $UPL$                   & \textbf{0.10730}         & \textbf{\uline{0.12886}} & \textbf{0.15073}         & \textbf{0.11787}         & \textbf{\uline{0.16416}} & \textbf{0.22510}         & \textbf{0.08923}         & \textbf{\uline{0.10375}} & \textbf{0.11751}          \\

\hline\hline
\end{tabular}

\caption{Ranking performance on Yahoo! R3 and Coat datasets. (The results of UPL with underline indicate p < 0.05 for one-tailed t-test with the best competitor. And $UBPR_{NClip}$' results with '-' indicate significant worse than UBPR (p<0.05).)}
\label{table1}
\end{table*}
\vspace{-0.1cm}

\subsection{Proposed Unbiased Pairwise Estimator} \label{section:PUPE}
First, the ideal pairwise risk function and the ranker learned through empirical risk minimization are defined in Equation \ref{eq:2} and Equation \ref{eq:3} respectively. With the ideal pairwise risk of Equation \ref{eq:2}, we try to minimize the overall loss for valid tuples of $(u, i, j)$.  
\vspace{-0.2cm}
\begin{equation} \label{eq:2}
 R_{ideal}(f)  =    \int_{}{} L(f(u,i), f(u,j)) dP(r_{u,i}=1, r_{u,j}=0) 
\end{equation}
\vspace{-0.8cm}

\begin{equation} \label{eq:3}
\hat{f}_{ideal}  =   \arg\min_{f}  ( \sum_{r_{u,i} =1,  r_{u,j}=0}  {L(f(u,i), f(u,j))} + \lambda Reg(f))
\end{equation}
\vspace{-0.3cm}

Then, according to equation \ref{eq:pair_relation}, we can prove the unbiased risk function defined on the implicit feedback pairs: 
\vspace{-0.25cm}
\begin{align} 
&\!\!\! \!  R_{UPL}(f) \nonumber  \\
&\!\!\! \! \!\!= \!\! \int_{}{}  \frac {L(f(u,i), f(u,j))} {\theta_{u,i} \theta_{u,j} } \! P(  o_{u,j}  \! =  \! 1|  c_{u,i}\!=\!1,c_{u,j}\!=\!0 ) dP( c_{u,i} \!\! = \!\!1,\! c_{u,j}\!\! = \!\!0 ) \! \nonumber  \\
& \!\!\! \! \!\!= \!\! \int_{}{}  \frac {L(f(u,i), f(u,j))} {\theta_{u,i} \theta_{u,j} }  dP(c_{u,i}\!\!=\!\!1,\! c_{u,j}\!\!=\!\!0, \!o_{u,i}\!\!=\!\!1, \!o_{u,j}\!\!=\!\!1) \! \nonumber  \\
&\!\!\! \! \!\!= \!\! \int_{}{}  \frac {L(f(u,i), f(u,j))} {\theta_{u,i} \theta_{u,j} }  dP( r_{u,i} \!\!=\!\!1, \!r_{u,j}\!\!=\!\!0)P(o_{u,i}\!\!=\!\!1)P(o_{u,j}\!\!=\!\!1) \!  \nonumber  \\
&\!\!\! \! \!\!= \!\! \int_{}{}  {L(f(u,i), f(u,j))} dP(r_{u,i}=1, r_{u,j}=0) \! = R_{ideal}(f)
\end{align}
As $R_{UPL}$ is an unbiased estimator of $R_{ideal}$, we can learn an unbiased ranker through minimizing the corresponding empirical loss function. Moreover, given independent exposure probabilities between different items, we can see $P(  o_{u,j}  \! =  \! 1|  c_{u,i}\!=\!1,c_{u,j}\!=\!0 )=P(  o_{u,j}  \! =  \! 1|  c_{u,j}\!=\!0 )= \theta_{u,j}^-$. All in all, the unbiased ranker can be learned based on Equation \ref{eq:6}, where $Reg(f)$ and $\lambda$ are the regularization loss and loss weight respectively. In contrast to UBPR, the empirical risk is only dependent on the valid pairs of implicit feedback, for which $c_{u,i}$ is greater than $c_{u,j}$.
\vspace{-0.1cm}
\begin{align} \label{eq:6}
& \!\!\!\!\!\hat{f}_{UPL}  =   \arg\min_{f}  (\sum_{c_{u,i}=1, c_{u,j}=0}  {\! \frac {L(f(u,i) f(u,j)) \theta_{u,j}^-} {\theta_{u,i} \theta_{u,j} }} +  \lambda Reg(f))\!\!  \nonumber \\
& \!\!\!\!\!=\arg\min_{f}  (\sum_{c_{u,i}=1, c_{u,j}=0} { \! \frac {L(f(u,i), f(u,j)) (1-\gamma_{u,j}) } {\theta_{u,i} (1 - \theta_{u,j}\gamma_{u,j})}}  + \lambda Reg(f) )\!\! 
\end{align}
\vspace{-0.1cm}
First, we can see the risk function cannot take negative values. Furthermore, we can derive its variance as below. Because of contrained space, we omit the detailed proof here. 
\vspace{-0.1cm}
\begin{align} \label{eq:7}
&\!\!\!\!\!\!VAR_{UPL}=\sum_{\substack{c_{u,i}=1 \\ c_{u,j}=0}}\frac{(\frac{1}{\theta_{u,i}}-\gamma_{u,i})\gamma_{u,i}(1-\gamma_{u,j})^2 L_{u,i,j}^2}{(1-\theta_{u,j}\gamma_{u,j})^2} \nonumber \\
&+\sum_{\substack{c_{u,i}=1, c_{u,j}=0\\ c_{u,k}=0, k \neq j}}\frac{(\frac{1}{\theta_{u,i}}-\gamma_{u,i})\gamma_{u,i}(1-\gamma_{u,j})(1-\gamma_{u,k})L_{u,i,j}L_{u,i,k}}{(1-\theta_{u,j}\gamma_{u,j})(1-\theta_{u,k}\gamma_{u,k})}
\end{align}
\vspace{-0.1cm}
where $L_{u,i,j}$ stands for $L(f(u,i) f(u,j))$. Since $(1-\theta_{u,j}\gamma_{u,j})$ is usually far away from 0 even for the most popular items and other factors, including $\gamma_{u,i}$ and $(1-\gamma_{u,j})$, are bounded between 0 and 1, the variance of Equation \ref{eq:7} largely depends on the inverse of $\theta_{u,i}$. In comparison, the variance of UBPR \cite{saito2020unbiased2} is impacted by the inverse of the product of $\theta_{u,i}$ and $\theta_{u,j}$, thereby producing a much greater variance. As a result, we do not adopt any further variance reduction techniques on top of the unbiased pairwise learning estimator.

Interestingly, if $\theta_{u,j}^-$ is approximated with $\theta_{u,j}$ in equation \ref{eq:6}, we can derive exactly the same estimator practically applied in UBPR after clipping all of the negative values. From this perspective, it is pretty obvious that the UBPR algorithm is not really unbiased in practice. To verify the necessity of variance control in UBPR, we will also remove the variance-bias trade-off step to test its performance in our experiments. 
\vspace{-0.2cm}

\subsection{Learning Algorithm}

\begin{algorithm}[!ht]  
\DontPrintSemicolon

  \KwData{implicit feedback data $\{c_{u,i}\};$ learned relevance from Rel-MF $\{\gamma_{u,i}\}$}  
  \KwInput{mini-batch size $M;$ regularization parameter $\lambda;$ learning\_rate $\alpha$}  
  \KwOutput{learned ranker $f$'s parameters $W$}

  Initialize parameters $W$ of Ranker $f$ \\
  Estimate propensity scores of $\{\theta_{u,i}\}$ \\
  \While{Not Convergence}
  {
   	 Sample one mini-batch data of size $M$ \\
     $Loss = \sum_{c_{u,i}=1, c_{u,j}=0} { \! \frac {L(f(u,i), f(u,j)) (1-\gamma_{u,j}) } {\theta_{u,i} (1 - \theta_{u,j}\gamma_{u,j})} } + \lambda Reg(f) $ \\
     Compute the gradient of $W$ based on $Loss$\\
     Update $W$ according to $\alpha$
  }
  Return $W$
\caption{Unbiased Pairwise Learning (\textbf{\textit{UPL}})} \label{alg:1}
\end{algorithm} 
\vspace{-0.2cm}

In this section, we formally describe the learning algorithm. First, the dependence on $\gamma_{u,j}$ is one problem of the above learning algorithm. Thanks to the work of Rel-MF \cite{saito2020unbiased}, theoretically, the Rel-MF model can predict unbiased relevance after convergence, which is also used as our input. Though the prediction results of Rel-MF may deviate, intuitively, we are likely to get better performance than UBPR since it uses a rather loose approximation for $\theta_{u,j}^-$. Then, the ranker $f$ is trained by optimizing the empirical risk function of Equation \ref{eq:6}. The whole learning procedure is illustrated in Algorithm \ref{alg:1}. As the model training is of a general process, it can be combined with any pairwise algorithms, such as \cite{rendle2012bpr,song2018neural,kim2019dual},  

\vspace{-0.2cm}

\section{Experiments}

\begin{table*}
\centering

\begin{tabular}{lllll|lll|lll} 
\hline
\multirow{2}{*}{Subsets}          & \multirow{2}{*}{Methods} & \multicolumn{3}{c|}{\textbf{\textit{DCG@K}}}                                   & \multicolumn{3}{c|}{\textbf{\textit{Recall@K}}}                                & \multicolumn{3}{c}{\textbf{\textit{MAP@K}}}                                     \\ 
\cline{3-11}
                                  &                          & \textit{K=3}             & \textit{K=5}             & \textit{K=8}             & \textit{K=3}             & \textit{K=5}             & \textit{K=8}             & \textit{K=3}             & \textit{K=5}             & \textit{K=8}              \\ 
\hline\hline
\multirow{6}{*}{Cold-start users} & WMF                      & 0.08406                  & 0.09618                  & 0.10773                  & 0.08984                  & 0.11584                  & 0.14790                  & 0.07002                  & 0.07791                  & 0.08465                   \\
                                  & Rel-MF                   & 0.08729                  & 0.09944                  & 0.11058                  & 0.09337                  & 0.11943                  & 0.15038                  & 0.07290                  & 0.08077                  & 0.08737                   \\
                                  & MF-DU                    & 0.08701                  & 0.09958                  & 0.11058                  & 0.09306                  & 0.11992                  & 0.15046                  & 0.07305                  & 0.08108                  & 0.08766                   \\
                                  & BPR                      & 0.09209                  & 0.10442                  & 0.11450                  & 0.09892                  & 0.12501                  & 0.15307                  & 0.07647                  & 0.08442                  & 0.09055                   \\
                                  & UBPR                     & 0.09560                  & 0.10880                  & 0.11792                  & 0.10230                  & 0.13035                  & 0.15537                  & 0.08061                  & 0.08942                  & 0.09507                   \\
                                  & UPL                   & \textbf{\uline{0.09689}} & \textbf{\uline{0.11005}} & \textbf{\uline{0.11905}} & \textbf{\uline{0.10345}} & \textbf{\uline{0.13139}} & \textbf{\uline{0.15626}} & \textbf{\uline{0.08278}} & \textbf{\uline{0.09159}} & \textbf{\uline{0.09711}}  \\ 
\hline
\multirow{6}{*}{Rare items}       & WMF                      & 0.04227                  & 0.05324                  & 0.06011                  & 0.04711                  & 0.07057                  & 0.08935                  & 0.03214                  & 0.03890                  & 0.04291                   \\
                                  & Rel-MF                   & 0.04454                  & 0.05484                  & 0.06141                  & 0.04940                  & 0.07147                  & 0.08943                  & 0.03429                  & 0.04064                  & 0.04456                   \\
                                  & MF-DU                    & 0.04545                  & 0.05565                  & 0.06199                  & 0.05032                  & 0.07210                  & 0.08940                  & 0.03520                  & 0.04156                  & 0.04532                   \\
                                  & BPR                      & 0.04720                  & 0.05690                  & 0.06287                  & 0.05208                  & 0.07283                  & 0.08917                  & 0.03713                  & 0.04311                  & 0.04672                   \\
                                  & UBPR                     & 0.05268                  & 0.06160                  & 0.06639                  & 0.05797                  & 0.07694                  & \textbf{0.08995}         & 0.04122                  & 0.04691                  & 0.04998                   \\
                                  & UPL                   & \textbf{\uline{0.05483}} & \textbf{\uline{0.06336}} & \textbf{\uline{0.06761}} & \textbf{\uline{0.06016}} & \textbf{\uline{0.07826}} & 0.08981                  & \textbf{\uline{0.04366}} & \textbf{\uline{0.04923}} & \textbf{\uline{0.05198}}  \\
\hline
\end{tabular}

\caption{Ranking performance for cold-start users and rare items on Yahoo! R3 (The results of UPL with underline indicate p < 0.05 for one-tailed t-test with the best competitor.)}
\label{table2}
\end{table*}
In this section, we demonstrate the effectiveness of the proposed \textbf{\textit{UPL}} algorithm. We open all of the related
source code on Github\footnote{https://github.com/renyi533/unbiased-pairwise-rec/tree/main}.
\vspace{-0.2cm}

\subsection{Experimental Setup}
\vspace{-0.1cm}

\subsubsection{Datasets}
We used Yahoo! R3\footnote{http://webscope.sandbox.yahoo.com/} and Coat\footnote{https://www.cs.cornell.edu/schnabts/mnar/}. 
They are selected because we can leverage the explicit feedback data with five-star ratings for relevance evaluation. Morevoer, they include both Missing-Not-At-Random training data and Missing-Completely-At-Random test data. First, we transform five-star ratings into relevance probability with $\gamma_{u,i}=\epsilon+(1-\epsilon)\frac{2^r-1}{2^{r_{max}}-1}$. We apply $\epsilon=0.1$ for training while $\epsilon=0$ for testing. Second, we sample binary relevance variable with $r_{u,i}=Bern(\gamma_{u,i})$. Third, we define the exposure variable $o_{u,i}$ based on whether user $u$ rated item $i$. Finally, we compute $c_{u,i}=o_{u,i}r_{u,i}$, which is the observed implicit feedback.
\vspace{-0.2cm}

\subsubsection{Evaluation Metrics}
We adopt three widely used implicit recommendation evaluation metrics – DCG@k (Discounted Cumulative Gain), Recall@k and MAP@k (Mean Average Precision). The detailed definition can be found in \cite{saito2020unbiased,saito2020unbiased2}. We report results with k = 3, 5, 8.
\vspace{-0.2cm}

\subsubsection{Models}
We tested pointwise algorithms of WMF \cite{hu2008collaborative}, Rel-MF \cite{saito2020unbiased}, MF-DU \cite{lee2021dual} and pairwise algorithms of BPR \cite{rendle2012bpr} and UBPR \cite{saito2020unbiased2} as baselines. For UPL, we combined with the BPR \cite{rendle2012bpr} algorithm. For UBPR, we also included the variant of removing the variance-bias trade-off step, named $UBPR_{NClip}$, to study the impact of its variance. We followed the definition of \cite{saito2020unbiased2} for rare items (less than 100 clicks in the training set) and cold-start users (less than six clicks in the training set) to study the performance of different algorithms in these two harder tasks.

Following \cite{lee2021dual}, for MF-DU, we separately computed the propensity scores for clicked and non-clicked items respectively as:
\vspace{-0.05cm}
\begin{equation}
\theta_{*,i}^{Click}=(\frac{n_i}{max_{i \in I}n_i})^{0.5} \;\;\;\; \theta_{*,i}^{NClick}=(1-\frac{n_i}{max_{i \in I}n_i})^{0.5}
\end{equation}
\vspace{-0.05cm}
where $n_i$ means the number of interactions to the item $i$ by all users. While for Rel-MF, UBPR and UPL, following \cite{saito2020unbiased,saito2020unbiased2}, we used $\theta_{*,i}^{Click}$ as the propensity scores for all of the items.

To be Consistent with \cite{saito2020unbiased2}, we randomly sampled 10\% of the training data as validation data to tune the user-item latent factors within [100, 300], L2 regularization hyperparameter within [$10^{-7},10^{-3}$], and the non-negative estimator's clipping threshold for UBPR in [-10, 0]. All models are implemented with tensorflow \cite{abadi2016tensorflow} and optimized using the Adam optimizer \cite{kingma2014adam} with an initial learning rate of 0.001 and a mini-batch size of 256. 

For each model, we performed 50 runs on both datasets to report the experiment results.  
\vspace{-0.4cm}
\subsection{Experiment Results}
\vspace{-0.1cm}
\subsubsection{Overall Offline Results}
For the overall ranking performance on both the Yahoo! R3 and Coat datasets, the results of all methods are presented in Table \ref{table1}. For both datasets, our UPL algorithm outperforms the other baseline methods under all settings. For the results of UPL, the underline symbol indicates a significant gain (p<0.05 for one tailed t-test) over the best competitor algorithm. For all the results of Yahoo! R3, UPL achieves significant performance gain. Though the results of Coat are of more variance, UPL still showes significant performance gain for the metrics of DCG@5, Recall@5 and MAP@5. These results support that our strategy of performing true unbiased pairwise learning is effective in improving the ranking performances. 

In addition, the pairwise algorithms show better ranking performance than the pointwise ones. For instance, BPR achieves even better results than the debiased pointwise algorithms in Yahoo! R3 datasets. And UBPR is the second best method for almost all settings. 

Finally, $UBPR_{NClip}$ is of worse performance than UBPR, especially for Yahoo R3 dataset, which verifies the necessity to perform external variance control for UBPR. Moreover, though the loss clipping threshold for UBPR is tuned between [-10, 0], the best performance is always derived with a value very near to 0, which is exactly one approximation of UPL by approximating $\theta_{u,j}^-$ with $\theta_{u,j}$ as described in section \ref{section:PUPE}.  

In summary, the proposed approach in this paper is helpful for addressing the positive-unlabeled and Missing-Not-At-Random issues, which constitute the main challenges for recommender model learning from implicit feedback data. 
\vspace{-0.3cm}

\subsubsection{Offline Results for Rare Items and Cold-start Users}
Considering the importance of recommendation for rare items and cold-start users in real-world scenarios, we added the experiment results on Yahoo R3! datasets for rare items and cold-start users in table \ref{table2}.
Except the metric of Recall@8 for rare items,  UPL achieves significant gain over the most competitive baselines. And for Recall@8 of rare items, UBPR does not show significant gain over UPL. Overall, we can draw the conclusion that the proposed method is a promising choice for constructing recommender systems based on implicit feedback. 
\vspace{-0.3cm}
\subsubsection{Online A/B Testing}
We verified the effectiveness of our proposed algorithm by replacing the existing BPR \cite{rendle2012bpr} model with the UPL model in the matching stage of a large-scale content recommender system at Tencent. There were tens of recall models deployed in the matching stage. The union of the outputs of these models was used as the input to the ranking stage. And the BPR model used to be the most important recall model by accounting for about 30\% displayed items. The seven days A/B testing showed a significant gain of 0.7\% for our main online metric of app stay time per person. Moreover, we classified the items to long-tail (<3,000 impressions), hot (500,000+) and others based on their exposure count before the A/B tesing. For long-tail and other contents, we observed more exposure by 1.31\% and 1.74\% respectively. While the impressions for hot items decreased 2.43\%. 
\vspace{-0.3cm}
\section{Conclusion}
\vspace{-0.15cm}
In this paper, we propose an unbiased pairwise learning method, named UPL, for recommender systems based on implicit feedback. We theoretically prove the unbiasedness of UPL and its advantage of lower variance over the existing method of UBPR \cite{saito2020unbiased2}. Extensive offline experiments and online A/B testing help to verify UPL's effectiveness at tackling the positive-unlabeled and Missing-Not-At-Random challenge for recommender model learning from the biased implicit feedback. Both the theoretical and empirical results suggest the proposed method is a competent candidate. 



\bibliographystyle{ACM-Reference-Format}
\balance
\bibliography{references}


\begin{thebibliography}{29}


\ifx \showCODEN    \undefined \def \showCODEN     #1{\unskip}     \fi
\ifx \showDOI      \undefined \def \showDOI       #1{#1}\fi
\ifx \showISBNx    \undefined \def \showISBNx     #1{\unskip}     \fi
\ifx \showISBNxiii \undefined \def \showISBNxiii  #1{\unskip}     \fi
\ifx \showISSN     \undefined \def \showISSN      #1{\unskip}     \fi
\ifx \showLCCN     \undefined \def \showLCCN      #1{\unskip}     \fi
\ifx \shownote     \undefined \def \shownote      #1{#1}          \fi
\ifx \showarticletitle \undefined \def \showarticletitle #1{#1}   \fi
\ifx \showURL      \undefined \def \showURL       {\relax}        \fi
\providecommand\bibfield[2]{#2}
\providecommand\bibinfo[2]{#2}
\providecommand\natexlab[1]{#1}
\providecommand\showeprint[2][]{arXiv:#2}

\bibitem[Abadi et~al\mbox{.}(2016)]%
        {abadi2016tensorflow}
\bibfield{author}{\bibinfo{person}{Mart{\'\i}n Abadi}, \bibinfo{person}{Paul
  Barham}, \bibinfo{person}{Jianmin Chen}, \bibinfo{person}{Zhifeng Chen},
  \bibinfo{person}{Andy Davis}, \bibinfo{person}{Jeffrey Dean},
  \bibinfo{person}{Matthieu Devin}, \bibinfo{person}{Sanjay Ghemawat},
  \bibinfo{person}{Geoffrey Irving}, \bibinfo{person}{Michael Isard},
  {et~al\mbox{.}}} \bibinfo{year}{2016}\natexlab{}.
\newblock \showarticletitle{$\{$TensorFlow$\}$: A System for
  $\{$Large-Scale$\}$ Machine Learning}. In \bibinfo{booktitle}{\emph{12th
  USENIX symposium on operating systems design and implementation (OSDI 16)}}.
  \bibinfo{pages}{265--283}.
\newblock


\bibitem[Agarwal et~al\mbox{.}(2019)]%
        {agarwal2019general}
\bibfield{author}{\bibinfo{person}{Aman Agarwal}, \bibinfo{person}{Kenta
  Takatsu}, \bibinfo{person}{Ivan Zaitsev}, {and} \bibinfo{person}{Thorsten
  Joachims}.} \bibinfo{year}{2019}\natexlab{}.
\newblock \showarticletitle{A general framework for counterfactual
  learning-to-rank}. In \bibinfo{booktitle}{\emph{Proceedings of the 42nd
  International ACM SIGIR Conference on Research and Development in Information
  Retrieval}}. \bibinfo{pages}{5--14}.
\newblock


\bibitem[Ai et~al\mbox{.}(2018)]%
        {ai2018unbiased}
\bibfield{author}{\bibinfo{person}{Qingyao Ai}, \bibinfo{person}{Keping Bi},
  \bibinfo{person}{Cheng Luo}, \bibinfo{person}{Jiafeng Guo}, {and}
  \bibinfo{person}{W~Bruce Croft}.} \bibinfo{year}{2018}\natexlab{}.
\newblock \showarticletitle{Unbiased learning to rank with unbiased propensity
  estimation}. In \bibinfo{booktitle}{\emph{The 41st International ACM SIGIR
  Conference on Research \& Development in Information Retrieval}}.
  \bibinfo{pages}{385--394}.
\newblock


\bibitem[Bekker et~al\mbox{.}(2019)]%
        {bekker2019beyond}
\bibfield{author}{\bibinfo{person}{Jessa Bekker}, \bibinfo{person}{Pieter
  Robberechts}, {and} \bibinfo{person}{Jesse Davis}.}
  \bibinfo{year}{2019}\natexlab{}.
\newblock \showarticletitle{Beyond the selected completely at random assumption
  for learning from positive and unlabeled data}. In
  \bibinfo{booktitle}{\emph{Joint European Conference on Machine Learning and
  Knowledge Discovery in Databases}}. Springer, \bibinfo{pages}{71--85}.
\newblock


\bibitem[Chen et~al\mbox{.}(2020)]%
        {chen2020bias}
\bibfield{author}{\bibinfo{person}{Jiawei Chen}, \bibinfo{person}{Hande Dong},
  \bibinfo{person}{Xiang Wang}, \bibinfo{person}{Fuli Feng},
  \bibinfo{person}{Meng Wang}, {and} \bibinfo{person}{Xiangnan He}.}
  \bibinfo{year}{2020}\natexlab{}.
\newblock \showarticletitle{Bias and debias in recommender system: A survey and
  future directions}.
\newblock \bibinfo{journal}{\emph{arXiv preprint arXiv:2010.03240}}
  (\bibinfo{year}{2020}).
\newblock


\bibitem[Elkan and Noto(2008)]%
        {elkan2008learning}
\bibfield{author}{\bibinfo{person}{Charles Elkan} {and} \bibinfo{person}{Keith
  Noto}.} \bibinfo{year}{2008}\natexlab{}.
\newblock \showarticletitle{Learning classifiers from only positive and
  unlabeled data}. In \bibinfo{booktitle}{\emph{Proceedings of the 14th ACM
  SIGKDD international conference on Knowledge discovery and data mining}}.
  \bibinfo{pages}{213--220}.
\newblock


\bibitem[Hu et~al\mbox{.}(2008)]%
        {hu2008collaborative}
\bibfield{author}{\bibinfo{person}{Yifan Hu}, \bibinfo{person}{Yehuda Koren},
  {and} \bibinfo{person}{Chris Volinsky}.} \bibinfo{year}{2008}\natexlab{}.
\newblock \showarticletitle{Collaborative filtering for implicit feedback
  datasets}. In \bibinfo{booktitle}{\emph{2008 Eighth IEEE international
  conference on data mining}}. Ieee, \bibinfo{pages}{263--272}.
\newblock


\bibitem[Hu et~al\mbox{.}(2019)]%
        {hu2019unbiased}
\bibfield{author}{\bibinfo{person}{Ziniu Hu}, \bibinfo{person}{Yang Wang},
  \bibinfo{person}{Qu Peng}, {and} \bibinfo{person}{Hang Li}.}
  \bibinfo{year}{2019}\natexlab{}.
\newblock \showarticletitle{Unbiased lambdamart: an unbiased pairwise
  learning-to-rank algorithm}. In \bibinfo{booktitle}{\emph{The World Wide Web
  Conference}}. \bibinfo{pages}{2830--2836}.
\newblock


\bibitem[Joachims et~al\mbox{.}(2017)]%
        {joachims2017unbiased}
\bibfield{author}{\bibinfo{person}{Thorsten Joachims}, \bibinfo{person}{Adith
  Swaminathan}, {and} \bibinfo{person}{Tobias Schnabel}.}
  \bibinfo{year}{2017}\natexlab{}.
\newblock \showarticletitle{Unbiased learning-to-rank with biased feedback}. In
  \bibinfo{booktitle}{\emph{Proceedings of the Tenth ACM International
  Conference on Web Search and Data Mining}}. \bibinfo{pages}{781--789}.
\newblock


\bibitem[Johnson(2014)]%
        {johnson2014logistic}
\bibfield{author}{\bibinfo{person}{Christopher~C Johnson}.}
  \bibinfo{year}{2014}\natexlab{}.
\newblock \showarticletitle{Logistic matrix factorization for implicit feedback
  data}.
\newblock \bibinfo{journal}{\emph{Advances in Neural Information Processing
  Systems}} \bibinfo{volume}{27}, \bibinfo{number}{78} (\bibinfo{year}{2014}),
  \bibinfo{pages}{1--9}.
\newblock


\bibitem[Kim et~al\mbox{.}(2019)]%
        {kim2019dual}
\bibfield{author}{\bibinfo{person}{Seunghyeon Kim}, \bibinfo{person}{Jongwuk
  Lee}, {and} \bibinfo{person}{Hyunjung Shim}.}
  \bibinfo{year}{2019}\natexlab{}.
\newblock \showarticletitle{Dual neural personalized ranking}. In
  \bibinfo{booktitle}{\emph{The World Wide Web Conference}}.
  \bibinfo{pages}{863--873}.
\newblock


\bibitem[Kingma and Ba(2014)]%
        {kingma2014adam}
\bibfield{author}{\bibinfo{person}{Diederik~P Kingma} {and}
  \bibinfo{person}{Jimmy Ba}.} \bibinfo{year}{2014}\natexlab{}.
\newblock \showarticletitle{Adam: A method for stochastic optimization}.
\newblock \bibinfo{journal}{\emph{arXiv preprint arXiv:1412.6980}}
  (\bibinfo{year}{2014}).
\newblock


\bibitem[Kiryo et~al\mbox{.}(2017)]%
        {kiryo2017positive}
\bibfield{author}{\bibinfo{person}{Ryuichi Kiryo}, \bibinfo{person}{Gang Niu},
  \bibinfo{person}{Marthinus~C Du~Plessis}, {and} \bibinfo{person}{Masashi
  Sugiyama}.} \bibinfo{year}{2017}\natexlab{}.
\newblock \showarticletitle{Positive-unlabeled learning with non-negative risk
  estimator}.
\newblock \bibinfo{journal}{\emph{Advances in neural information processing
  systems}}  \bibinfo{volume}{30} (\bibinfo{year}{2017}).
\newblock


\bibitem[Koren et~al\mbox{.}(2009)]%
        {koren2009matrix}
\bibfield{author}{\bibinfo{person}{Yehuda Koren}, \bibinfo{person}{Robert
  Bell}, {and} \bibinfo{person}{Chris Volinsky}.}
  \bibinfo{year}{2009}\natexlab{}.
\newblock \showarticletitle{Matrix factorization techniques for recommender
  systems}.
\newblock \bibinfo{journal}{\emph{Computer}} \bibinfo{volume}{42},
  \bibinfo{number}{8} (\bibinfo{year}{2009}), \bibinfo{pages}{30--37}.
\newblock


\bibitem[Lee et~al\mbox{.}(2021)]%
        {lee2021dual}
\bibfield{author}{\bibinfo{person}{Jae-woong Lee}, \bibinfo{person}{Seongmin
  Park}, {and} \bibinfo{person}{Jongwuk Lee}.} \bibinfo{year}{2021}\natexlab{}.
\newblock \showarticletitle{Dual Unbiased Recommender Learning for Implicit
  Feedback}. In \bibinfo{booktitle}{\emph{Proceedings of the 44th International
  ACM SIGIR Conference on Research and Development in Information Retrieval}}.
  \bibinfo{pages}{1647--1651}.
\newblock


\bibitem[Liang et~al\mbox{.}(2016)]%
        {liang2016modeling}
\bibfield{author}{\bibinfo{person}{Dawen Liang}, \bibinfo{person}{Laurent
  Charlin}, \bibinfo{person}{James McInerney}, {and} \bibinfo{person}{David~M
  Blei}.} \bibinfo{year}{2016}\natexlab{}.
\newblock \showarticletitle{Modeling user exposure in recommendation}. In
  \bibinfo{booktitle}{\emph{Proceedings of the 25th international conference on
  World Wide Web}}. \bibinfo{pages}{951--961}.
\newblock


\bibitem[Liu et~al\mbox{.}(2010)]%
        {liu2010personalized}
\bibfield{author}{\bibinfo{person}{Jiahui Liu}, \bibinfo{person}{Peter Dolan},
  {and} \bibinfo{person}{Elin~R{\o}nby Pedersen}.}
  \bibinfo{year}{2010}\natexlab{}.
\newblock \showarticletitle{Personalized news recommendation based on click
  behavior}. In \bibinfo{booktitle}{\emph{Proceedings of the 15th international
  conference on Intelligent user interfaces}}. \bibinfo{pages}{31--40}.
\newblock


\bibitem[Ren et~al\mbox{.}(2022)]%
        {ren2022unbiased}
\bibfield{author}{\bibinfo{person}{Yi Ren}, \bibinfo{person}{Hongyan Tang},
  {and} \bibinfo{person}{Siwen Zhu}.} \bibinfo{year}{2022}\natexlab{}.
\newblock \showarticletitle{Unbiased Learning to Rank with Biased Continuous
  Feedback}. In \bibinfo{booktitle}{\emph{Proceedings of the 31st ACM
  International Conference on Information \& Knowledge Management}}.
  \bibinfo{pages}{1716--1725}.
\newblock


\bibitem[Rendle et~al\mbox{.}(2012)]%
        {rendle2012bpr}
\bibfield{author}{\bibinfo{person}{Steffen Rendle}, \bibinfo{person}{Christoph
  Freudenthaler}, \bibinfo{person}{Zeno Gantner}, {and} \bibinfo{person}{Lars
  Schmidt-Thieme}.} \bibinfo{year}{2012}\natexlab{}.
\newblock \showarticletitle{BPR: Bayesian personalized ranking from implicit
  feedback}.
\newblock \bibinfo{journal}{\emph{arXiv preprint arXiv:1205.2618}}
  (\bibinfo{year}{2012}).
\newblock


\bibitem[Rosenbaum and Rubin(1983)]%
        {rosenbaum1983central}
\bibfield{author}{\bibinfo{person}{Paul~R Rosenbaum} {and}
  \bibinfo{person}{Donald~B Rubin}.} \bibinfo{year}{1983}\natexlab{}.
\newblock \showarticletitle{The central role of the propensity score in
  observational studies for causal effects}.
\newblock \bibinfo{journal}{\emph{Biometrika}} \bibinfo{volume}{70},
  \bibinfo{number}{1} (\bibinfo{year}{1983}), \bibinfo{pages}{41--55}.
\newblock


\bibitem[Saito(2020)]%
        {saito2020unbiased2}
\bibfield{author}{\bibinfo{person}{Yuta Saito}.}
  \bibinfo{year}{2020}\natexlab{}.
\newblock \showarticletitle{Unbiased Pairwise Learning from Biased Implicit
  Feedback}. In \bibinfo{booktitle}{\emph{Proceedings of the 2020 ACM SIGIR on
  International Conference on Theory of Information Retrieval}}.
  \bibinfo{pages}{5--12}.
\newblock


\bibitem[Saito et~al\mbox{.}(2020)]%
        {saito2020unbiased}
\bibfield{author}{\bibinfo{person}{Yuta Saito}, \bibinfo{person}{Suguru
  Yaginuma}, \bibinfo{person}{Yuta Nishino}, \bibinfo{person}{Hayato Sakata},
  {and} \bibinfo{person}{Kazuhide Nakata}.} \bibinfo{year}{2020}\natexlab{}.
\newblock \showarticletitle{Unbiased recommender learning from
  missing-not-at-random implicit feedback}. In
  \bibinfo{booktitle}{\emph{Proceedings of the 13th International Conference on
  Web Search and Data Mining}}. \bibinfo{pages}{501--509}.
\newblock


\bibitem[Schnabel et~al\mbox{.}(2016)]%
        {schnabel2016recommendations}
\bibfield{author}{\bibinfo{person}{Tobias Schnabel}, \bibinfo{person}{Adith
  Swaminathan}, \bibinfo{person}{Ashudeep Singh}, \bibinfo{person}{Navin
  Chandak}, {and} \bibinfo{person}{Thorsten Joachims}.}
  \bibinfo{year}{2016}\natexlab{}.
\newblock \showarticletitle{Recommendations as treatments: Debiasing learning
  and evaluation}. In \bibinfo{booktitle}{\emph{international conference on
  machine learning}}. PMLR, \bibinfo{pages}{1670--1679}.
\newblock


\bibitem[Song et~al\mbox{.}(2018)]%
        {song2018neural}
\bibfield{author}{\bibinfo{person}{Bo Song}, \bibinfo{person}{Xin Yang},
  \bibinfo{person}{Yi Cao}, {and} \bibinfo{person}{Congfu Xu}.}
  \bibinfo{year}{2018}\natexlab{}.
\newblock \showarticletitle{Neural collaborative ranking}. In
  \bibinfo{booktitle}{\emph{Proceedings of the 27th ACM International
  Conference on Information and Knowledge Management}}.
  \bibinfo{pages}{1353--1362}.
\newblock


\bibitem[Steck(2010)]%
        {steck2010training}
\bibfield{author}{\bibinfo{person}{Harald Steck}.}
  \bibinfo{year}{2010}\natexlab{}.
\newblock \showarticletitle{Training and testing of recommender systems on data
  missing not at random}. In \bibinfo{booktitle}{\emph{Proceedings of the 16th
  ACM SIGKDD international conference on Knowledge discovery and data mining}}.
  \bibinfo{pages}{713--722}.
\newblock


\bibitem[Swaminathan and Joachims(2015)]%
        {swaminathan2015self}
\bibfield{author}{\bibinfo{person}{Adith Swaminathan} {and}
  \bibinfo{person}{Thorsten Joachims}.} \bibinfo{year}{2015}\natexlab{}.
\newblock \showarticletitle{The self-normalized estimator for counterfactual
  learning}.
\newblock \bibinfo{journal}{\emph{advances in neural information processing
  systems}}  \bibinfo{volume}{28} (\bibinfo{year}{2015}).
\newblock


\bibitem[Wang et~al\mbox{.}(2018b)]%
        {wang2018collaborative}
\bibfield{author}{\bibinfo{person}{Menghan Wang}, \bibinfo{person}{Xiaolin
  Zheng}, \bibinfo{person}{Yang Yang}, {and} \bibinfo{person}{Kun Zhang}.}
  \bibinfo{year}{2018}\natexlab{b}.
\newblock \showarticletitle{Collaborative filtering with social exposure: A
  modular approach to social recommendation}. In
  \bibinfo{booktitle}{\emph{Proceedings of the AAAI Conference on Artificial
  Intelligence}}, Vol.~\bibinfo{volume}{32}.
\newblock


\bibitem[Wang et~al\mbox{.}(2018a)]%
        {wang2018position}
\bibfield{author}{\bibinfo{person}{Xuanhui Wang}, \bibinfo{person}{Nadav
  Golbandi}, \bibinfo{person}{Michael Bendersky}, \bibinfo{person}{Donald
  Metzler}, {and} \bibinfo{person}{Marc Najork}.}
  \bibinfo{year}{2018}\natexlab{a}.
\newblock \showarticletitle{Position bias estimation for unbiased learning to
  rank in personal search}. In \bibinfo{booktitle}{\emph{Proceedings of the
  Eleventh ACM International Conference on Web Search and Data Mining}}.
  \bibinfo{pages}{610--618}.
\newblock


\bibitem[Yang et~al\mbox{.}(2018)]%
        {yang2018unbiased}
\bibfield{author}{\bibinfo{person}{Longqi Yang}, \bibinfo{person}{Yin Cui},
  \bibinfo{person}{Yuan Xuan}, \bibinfo{person}{Chenyang Wang},
  \bibinfo{person}{Serge Belongie}, {and} \bibinfo{person}{Deborah Estrin}.}
  \bibinfo{year}{2018}\natexlab{}.
\newblock \showarticletitle{Unbiased offline recommender evaluation for
  missing-not-at-random implicit feedback}. In
  \bibinfo{booktitle}{\emph{Proceedings of the 12th ACM conference on
  recommender systems}}. \bibinfo{pages}{279--287}.
\newblock


\end{thebibliography}

\end{document}